\documentclass[fleqn,twoside]{article}
\usepackage{espcrc2}

\usepackage{graphicx}
\usepackage[figuresright]{rotating}

\newcommand{\AmS}{{\protect\the\textfont2
  A\kern-.1667em\lower.5ex\hbox{M}\kern-.125emS}}

\title{A partially quenched analysis of the $\eta-\eta'$ system
in $N_f = 2$ QCD}

\author{H. Neff\address{Center for Computational Science,
        Boston University,  
        3 Cummington Street, Boston MA02215, USA},
        Th. Lippert\address[WUPP]{Fachbereich Physik, Bergische Universit\"at
        Wuppertal,  D-42097 Wuppertal, Gaussstrasse,  Germany},
        J. Negele\address{Center for Theoretical Physics, 
        Laboratory for Nuclear Science and Physics, Massachusetts Institute of Technology, Cambridge MA02139, USA}
        and
        K. Schilling\addressmark[WUPP]}
       
\begin{document}

\begin{abstract}
  We report on a first, comprehensive partially quenched study of the
  $\eta-\eta'$ problem, based on SESAM configuration on a $16^3\times32$
  lattice at $\beta=5.6$ QCD with two (mass degenerate) active sea quark flavours, $N_f=2$.  By
  means of the spectral approximation of the two-loop (hairpin) diagrams, we
  find clear plateau formation in the effective masses which enables us both to
  determine   the $\eta$-$\eta'$ mass matrix and 
the $\alpha$-parameter in the
   effective chiral Lagrangian for the  flavour singlet sector,
  $\alpha = 0.028 \pm 0.013$.  \vspace{1pc}
\end{abstract}

\maketitle

\section{INTRODUCTION}
Flavour singlet pseudoscalar states bear witness to the topological structure
of the QCD vacuum and hence constitute objects of great physical interest. In
lattice gauge theory, considerable progress has been achieved
recently~\cite{us} in directly accessing such $\eta'$ like states from their
propagators in $N_f=2$ QCD with Wilson fermion action, based on applying
spectral approximations to two-loop correlators (hairpin diagrams).

\begin{figure}[t]
\vskip -1.3cm
\includegraphics[bb = 00 50 200 300,
width=.22\columnwidth,angle=270]{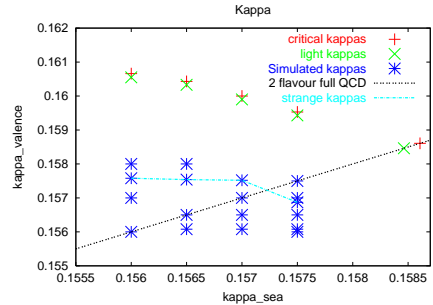}
\vskip 2cm
\caption{Our points of operation 
in the hopping parameter plane of valence/sea quarks.}
\label{fig:operational}
\end{figure}

In this note, we extend this work to the more realistic setting  of
partially quenched QCD with two active sea quark flavours, where valence and
sea quarks are allowed to carry different masses.  In our  comprehensive
study we  analyzed some 800 independent SESAM configurations on
$16^3\times 32$ lattices at $\beta = 5.6$~\cite{sesam}, with four different sea quark
masses, as indicated  in Fig.\ref{fig:operational}~\footnote{For further details of
  the analysis,   see Refs.~\cite{us,announce}.}.

\section{The partially quenched scenario}
Partial quenching in the flavour singlet sector of QCD is described in the
framework of chiral perturbation theory by  a nonsymmetric piece of  the
effective chiral Lagrangian~\cite{BG}
\begin{equation}
{\cal L}_0 =  + \frac{N_f}{2}\Big [{\mu^2} (\eta')^2 +
{ \alpha}
(\partial_{\mu}\eta')^2\Big ] \; .
\end{equation}

This expression introduces
two effective `low energy constants': the mass gap $\mu$ and 
an  interaction parameter  accounting  for momentum dependence, $\alpha$.
The Lagrangian translates into two-point correlation functions (propagators)
for the neutral pseudoscalar quark bilinears
\begin{equation}
\Phi_{ii}(x) = \bar{q}_i(x)\gamma_5 q_i(x)\; , 
\label{eq:lagr}
\end{equation}
where the index $i$ runs over the quark and pseudoquarks {\it d.o.f.}, the
latter being geared to act  as `determinant eaters' on the valence
quarks~\cite{BG}.

\begin{figure}[t]
\vspace{9pt}
\includegraphics[width=.7\columnwidth,angle=270]{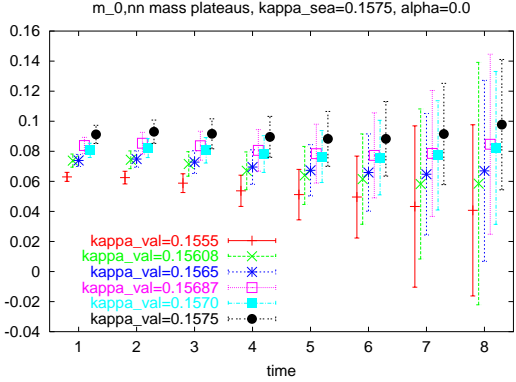}
\vskip -.6cm
\caption{First evidence  of plateau formation
  in a partially quenched setting, illustrated on the effective mass gap,
  $m_{0nn}(t)$, in the disconnected two loop correlator, $D_{nn}(t)$. The data refer
  to the lightest sea quark mass of SESAM, i.e. $\kappa_{sea} = .1575$, for
  various isosinglet valence quark masses.}
\label{fig:plateau_signal}
\end{figure}

In momentum space the disconnected part of the ps-correlators
has a compact form~\cite{GP} with the infamous double pole structure
\begin{equation}
D_{ij}(p) =  {\cal N} \frac{({ \mu^2} +  \alpha p^2)(p^2+
  M_d^2)}{(p^2+M_{ii}^2)(p^2+M_{jj}^2)(p^2+M_{\eta'}^2)}
\label{eq:form}
\end{equation}
The prefactor, ${\cal N} = 1/(1 + N_f \alpha )$ and the singlet mass, $
M^2_{\eta'} = (M^2_d + N_f \mu^2)/(1 + N_f \alpha )$ carry explicit
$\alpha$-dependencies. Rememeber that the octet masses ($M_{ii}$,$M_{jj}$, and
$M_d$) from valence and sea quarks, respectively, are readily computed from
connected diagrams on the lattice. Our present concern is  the determination of
the mass gap, $\mu$, and $\alpha$ from our lattice data. 

\begin{figure}[t]
\vspace{9pt}
\includegraphics[width=.7\columnwidth,angle=270]{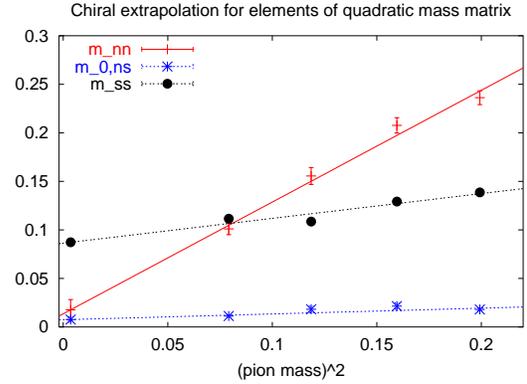}
\caption{Extrapolations of the quadratic mass matrix  elements of light ($n$) and strange ($s$)
 isosinglet  valence quarks   to QCD with {\it chiral} sea quarks (Eq.~\ref{eq:MM}), at $\alpha
 = 0$.}
\label{fig:chiral_signal}
\end{figure}

\section{Results}
As it has not been possible  to determine the  effective constant
$\alpha$ in the past we explore the potential of  our data for
achieving progress on this issue by applying
our spectral techniques~\cite{us} in two steps.
\subsection{$\alpha=0$ analysis}
In a first attempt, we choose $\alpha = 0$ and study effective mass gaps as
determined from the hairpin diagrams, $D_{ij}(t)$, by fitting, w.r.t.
$m_{0ij}(t) := \mu$ at each value of $t$, the zero-momentum Fourier transforms
of ansatz Eq.~\ref{eq:form} to our data.  In broken SU(3) with two active
quark flavours, we introduce two isosinglet pseudoscalars, designated by the
indices $n$ (for `light', nonstrange) and $s$ (for strange). In this manner,
we can study three types of hairpin diagrams, $D_{nn}$ , $D_{ns}$ , $D_{ss}$,
with the strange quark mass determined on the lattice from the kaon mass (for
$\kappa$ locations, see Fig.1).  As a result, we find satisfying  plateau formations in
$m_{0ij}(t)$ as illustrated in Fig.~\ref{fig:plateau_signal}.  In fact, the
numerical quality of our signals comfortably allows for   the consecutive extrapolations
{\it (i)} to light valence and then {\it (ii)} to light sea quark masses; the
chiral sea quark extrapolation is exhibited in Fig.~\ref{fig:chiral_signal}.

\begin{figure}[t]
\vskip -.4cm
\includegraphics[width=.7\columnwidth,angle=270]{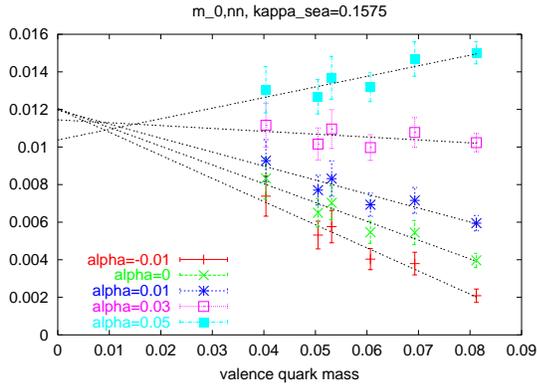}
\caption{Minimizing the  valence quark dependence of the mass gap by 
variation of   $\alpha$.}
\label{fig:minimize}
\end{figure}

From the consecutive chiral extrapolations, we obtain for the
quadratic mass matrix in the quark-flavour basis
\begin{eqnarray}
 {\cal M} := \left( 
\begin{array}{cc}
 m^2_{nn}  &  m^2_{0,ns}\\
 m^2_{0,sn}  &  m^2_{ss} 
\end{array}
\label{eq:MM}
\right) \; ,
\end{eqnarray}
where $m^2_{nn} := M^2_{nn} + m^2_{0nn}$.
With  the $\rho$-meson mass scale~\cite{sesam} we find
\begin{eqnarray}
{\cal M} =
 \left( 
\begin{array}{cc}
 (306 \pm 91)^2 &  \sqrt{2}(201 \pm 34)^2\\
 \sqrt{2}(201 \pm 34)^2  &  (680 \pm 15)^2 
\end{array}  
\right) \mbox{MeV}^2 \; .
\nonumber
\end{eqnarray}
Diagonalization  of $\cal M$ renders 
\begin{equation}
 M_{\eta} = 292 \pm 31 \; \mbox{MeV}, \quad M_{\eta'} = 686 \pm 31 \;
\mbox{MeV}.
\label{eq:values}
\end{equation}
These numbers from $N_f=2$ look promising: the individual mass values --
while being  low {\it w.r.t.} the real three flavour world -- 
yield a  mass splitting that  compares  very well to  phenomenology.

\subsection{First determination of $\alpha$}
We remark that the outcome of the analysis at this stage is not in accord with the
tree approximation to the Lagrangian, Eq.~\ref{eq:lagr}, as the latter predicts
the mass plateaus  to be {\it independent} of the valence quark mass, $m_v$, contrary
to the findings in Fig.~\ref{fig:plateau_signal}. Therefore, we 
repeated the above
mass plateau study on a set of nonvanishing  $\alpha$-parameters, in
an attempt to verify  the compliance of   our data with such independence: inspecting the plot
of $m_{0nn}$ vs. $m_v$ (Fig.~\ref{fig:minimize}) it becomes obvious  {\it (i)}
that $m_{0nn}(m_v)$ shows a simple, linear behaviour and {\it (ii)}  that $\alpha$
can indeed be adjusted  to produce a zero slope of $m_{0nn}(m_v)$~\footnote{The
  situation is less favourable for the gap values extracted from $D_{ns}\cite{announce}.$}!

\begin{figure}[t]
\vskip -.3cm
\includegraphics[width=.7\columnwidth,angle=270]{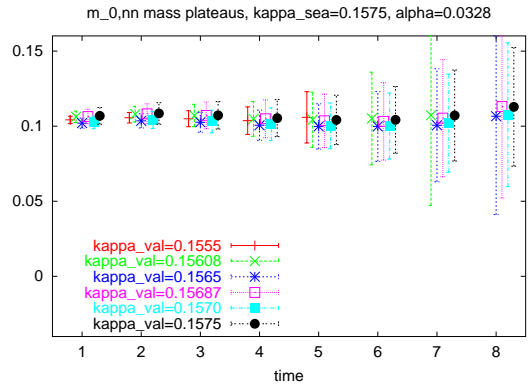}
\caption{Univerval plateau formation, at the lightest sea quak mass,
adjusting to $\alpha = 0.0328$.}
\label{fig:universal}
\end{figure}

In this way it is straightforward to find an optimal value, $\alpha_{opt}$,
that eliminates (for a given sea quark mass) the $m_v$-dependence from  $m_{0nn}$, 
resulting in a universal plateau level.
This collapse of data 
into  universal plateaus (of heights $\tilde{\mu}$)
is exemplified in Fig.~\ref{fig:universal}.
The comparision with the situation
encountered  with $\alpha = 0$ (Fig.~\ref{fig:plateau_signal}) provides clear evidence
for our sensitivity in  determining $\alpha_{opt}$.

 After chiral sea quark extrapolation we arrive at the values
\begin{equation}
\alpha = 0.028 \pm 0.013 \quad \mu = 203 \pm 34 \mbox{MeV} \; .
\end{equation}
Note that the numerical value for the mass gap {\it at the chiral point} is robust w.r.t. to the
two different approaches presented here. One might say that the allowance for   the $\alpha$ parameter
anticipates the chiral valence quark limits.

In conclusion: the application of spectral methods to the hairpin diagrams
allows rather detailed partially quenched studies in the flavour singlet
sector, with standard Wilson fermions. This provides a strong motivation to
extend this work to the case of  overlap fermions.

{\bf Acknowledgements} 
We thank G. Rossi and M. Golterman  for interesting discussions.
The project was supported by NERSC,  FZ-J\"ulich and 
EU (project HPRN-CT-2000-00145).

\end{document}